\documentclass[conference]{IEEEtran}

\usepackage{cite}
\usepackage{amsmath,amssymb,amsfonts}
\usepackage{algorithmic}
\usepackage{booktabs}
\usepackage{graphicx}
\usepackage{textcomp}
\usepackage{xcolor}
\def\BibTeX{{\rm B\kern-.05em{\sc i\kern-.025em b}\kern-.08em
    T\kern-.1667em\lower.7ex\hbox{E}\kern-.125emX}}

\begin{document}

\title{WiFi Sensing via Reservoir Computing}

\author{
\IEEEauthorblockN{
Heping Wang,
Zhongqin Wang,
J. Andrew Zhang
}
\IEEEauthorblockA{
School of Electrical and Data Engineering, University of Technology Sydney, NSW 2007,Australia
}

\IEEEauthorblockA{
heping.wang@student.uts.edu.au, \{zhongqin.wang, andrew.zhang\}@uts.edu.au
}
}

\maketitle

\begin{abstract}

Practical WiFi sensing must handle clock-asynchronous links, cross-domain variation, and post-deployment updating under the limited compute budget of access point (AP), router, and embedded Internet-of-Things platforms such as ESP-class devices. Reservoir computing (RC) is attractive in this setting because its temporal encoder can remain fixed while only a lightweight readout needs to be optimized and updated. To address these deployment challenges under tight compute budgets, we present ReWiS, a WiFi-sensing-oriented reservoir framework that transforms channel state information (CSI) into structured micro-Doppler streams with common, antenna-specific, and differential motion cues, encodes them with a graph-coupled reservoir, and adapts to a new domain after deployment by freezing the reservoir and fine-tuning only a compact readout with a few labeled target samples. On a large-scale WiFi sensing benchmark, ReWiS achieves 89.2\% in-domain macro-F1 and 82.0\% mean cross-domain macro-F1 with only a 0.72M trainable readout, improves to 88.5\% after lightweight post-deployment adaptation, and remains competitive with recent deep baselines evaluated under the same protocol, which achieve 87.5\%--89.2\% mean cross-domain macro-F1, while requiring lower optimization cost and lower CPU latency. These results indicate that ReWiS provides a practical reservoir-based design for deployable WiFi sensing, with further potential for low-power hardware realization.

\end{abstract}


\vspace{-0.8em}

\section{Introduction}

WiFi sensing is appealing because it can reuse commodity access points, routers for privacy-preserving perception without cameras or wearables. In practical deployment, however, the sensing stack must cope with clock asynchrony, strong cross-domain variation, and online update after installation, all under tight compute and memory budgets on AP/router/ESP-class hardware. Recent work has substantially improved signal-side synchronization, calibration, and sensing-oriented preprocessing for practical wireless links \cite{wu2024sensing,wang2023widfs}. Broader integrated sensing-and-communications studies have further clarified how these front-end issues can be handled within future wireless systems \cite{liu2022isac}. The more persistent bottleneck is now on the model side: how to keep WiFi sensing robust under domain shift while leaving the deployed model light enough for commodity devices.

Existing WiFi sensing models have evolved from handcrafted channel state information (CSI) statistics to learned motion-sensitive representations \cite{xie2024wiam}. Recent cross-domain systems have further pushed the field toward domain generalization and unseen-environment robustness \cite{wang2024airfi,li2025crossdomain}. Large-scale evaluation settings such as Widar 3.0\cite{zhang2022widar3} have also made WiFi sensing benchmarks more realistic and more challenging. These efforts have clearly advanced practical WiFi sensing and established strong baselines for cross-domain evaluation.

Three model-side challenges therefore remain central. First, post-deployment adaptation is still difficult because once a sensing node is installed, a small amount of target-domain supervision may become available, but re-optimizing the full temporal encoder is usually impractical on AP/router/ESP-class devices. Second, lightweight temporal modeling is still needed because stronger performance often comes from larger end-to-end backbones whose optimization and inference costs remain heavy for these platforms. Third, structured multichannel representation is still insufficient because existing models often use either raw CSI measurements or a single derived measurement as input, without explicitly separating common motion components, antenna-specific sensing views, and cross-antenna differential cues.

\begin{figure*}[!t]
    \centering
    \includegraphics[width=\textwidth]{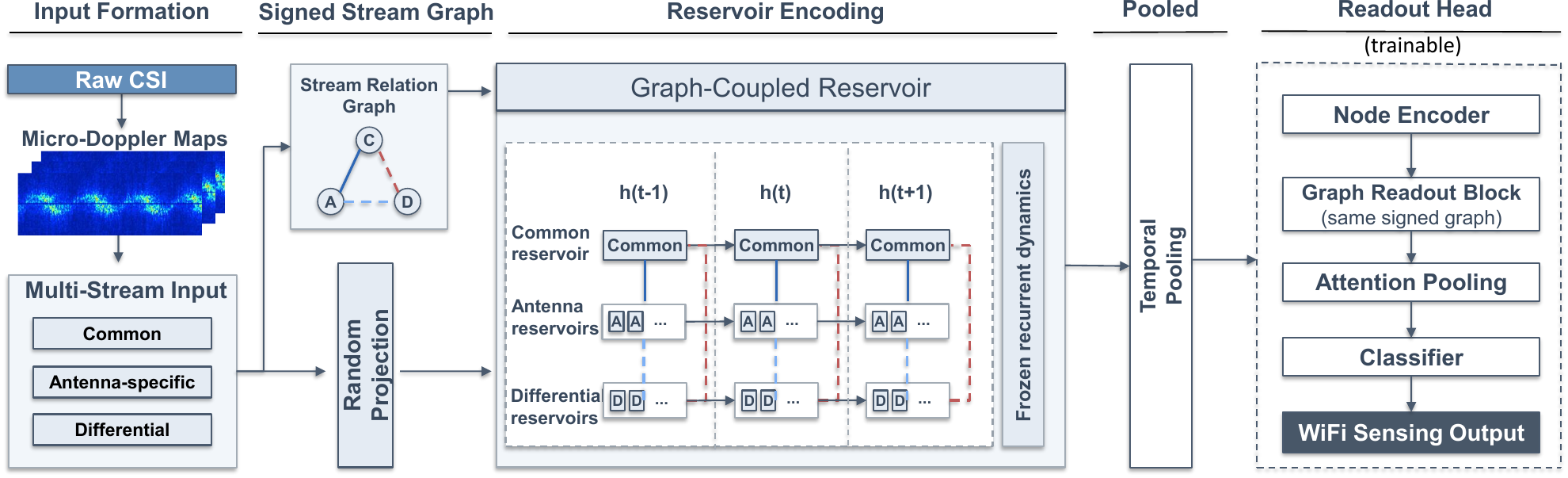}
\caption{Overview of ReWiS. Preprocessed CSI is transformed into antenna-wise micro-Doppler views, reorganized into structured common, antenna-specific, and differential streams, encoded by a fixed graph-coupled reservoir, and classified by a lightweight readout that is later adapted with a compact output head.}
    \label{fig:rewig_overview}
\end{figure*}

Reservoir computing (RC) is a lightweight temporal modeling technique in which the recurrent dynamics remain fixed and only the readout is trained. This property is particularly appealing for WiFi sensing, where CSI-derived wireless observations are dynamic time series, and WiFi sensing depends on the nonlinear temporal evolution of channel features. Compared with conventional deep temporal backbones that optimize all encoder weights end to end, RC keeps the recurrent dynamics fixed and trains only the readout, making it well matched to short-memory temporal encoding with lower optimization burden and to future WiFi sensing nodes where hardware reservoirs can serve as fixed front-end temporal encoders \cite{zhang2023survey}. Recent physical RC studies further suggest that lightweight hardware realizations are feasible beyond purely digital implementations \cite{allwood2023physicalrc,liang2024physicalrc}. These properties indicate that RC can potentially overcome the above challenges in practical WiFi sensing under domain variation. 

Motivated by this view, we propose ReWiS, a deployment-oriented RC framework for WiFi sensing. ReWiS introduces a sensing-motivated multistream micro-Doppler representation and signed stream graph, develops a graph-coupled reservoir encoder with fixed recurrent dynamics, and enables post-deployment adaptation through a lightweight readout head. This design directly addresses the above three challenges by introducing physically meaningful multichannel structure, reducing encoder-side optimization cost, and restricting post-deployment update to a small decision head. On Widar 3.0 \cite{zhang2022widar3}, which contains 271,013 samples from 22 gesture classes across 17 users, 3 rooms, 6 receiver positions, 5 face orientations, and 8 torso locations, ReWiS reaches 89.2\% in-domain macro-F1 and 82.0\% mean cross-domain macro-F1, and rises to 88.5\% when only a few target samples per class are available after deployment. Under the same evaluation protocol, the deep baselines report 87.5\%--89.2\% mean cross-domain macro-F1, providing a strong reference level for this benchmark. This deployment-oriented RC formulation also leaves room for future edge-side realization through hardware reservoirs integrated into practical WiFi sensing nodes.

The rest of the paper is organized as follows. Section~II details the sensing input layer, graph-coupled reservoir, and readout-only deployment adaptation; Section~III presents the evaluation protocol and baselines; Section~IV reports the results; Section~V concludes the paper.

\section{The Proposed Framework}

 As shown in Fig.~\ref{fig:rewig_overview}, ReWiS builds on RC but introduces three sensing-oriented design novelties. It first converts preprocessed CSI into structured micro-Doppler streams with common, antenna-specific, and differential motion cues, then encodes these streams with a graph-coupled reservoir guided by a signed stream graph, and performs prediction and post-deployment adaptation through a lightweight readout head. At a high level, the prediction pipeline can be written as
\begin{equation}
\widehat{\mathbf{y}}=f_{\mathrm{cls}}\!\left(\mathcal{G}\!\left(\mathcal{P}\!\left(\mathcal{R}(\mathcal{X};\mathbf{A})\right)\right)\right),
\end{equation}
where $\mathcal{X}$ denotes the structured WiFi-stream input set, $\mathbf{A}$ denotes the signed stream graph prior, $\mathcal{R}(\cdot)$ denotes graph-coupled reservoir encoding, $\mathcal{P}(\cdot)$ denotes temporal pooling, $\mathcal{G}(\cdot)$ denotes lightweight graph readout, and $f_{\mathrm{cls}}(\cdot)$ denotes the final classifier. We now describe these three parts in the same order, namely the sensing input layer, the fixed reservoir encoder, and the final lightweight output head.

\subsection{Structured Micro-Doppler Stream Construction}
\subsubsection{Preprocessing Motivation}
ReWiS does not send raw CSI tensors directly into the reservoir. Instead, it first converts each receive channel into a micro-Doppler map, following the preprocessing pipeline adapted from \cite{wang2025sisoisac}. Micro-Doppler preserves fine-grained motion-induced frequency modulation and is more informative for subtle human motion than generic channel magnitudes alone. Let $\mathbf{H}_{\mathrm{CSI}}[n]\in\mathbb{C}^{K\times M}$ denote the calibrated CSI matrix at packet index $n$, where $K$ is the number of active subcarriers and $M$ is the number of receive channels. Following the sensing pipeline used in our earlier WiFi modeling work \cite{wang2025sisoisac}, each antenna stream is first interpolated onto a dense subcarrier grid, then used to reconstruct the dominant static response in the delay domain while retaining the same timing offset and carrier-frequency offset as the original CSI, and finally conjugate-multiplied with the original CSI to suppress shared clock-asynchronous distortion and obtain a motion-sensitive sequence. For the $m$-th receive channel, we write the corresponding subcarrier response as $\mathbf{h}_m[n]\in\mathbb{C}^{K}$ and define the range-focused motion feature as
\begin{equation}
\mathbf{c}_m[n]=\widetilde{\mathbf{h}}_m[n]\odot \widetilde{\mathbf{h}}_{m,\mathrm{ref}}^{*}[n],
\end{equation}
where $\widetilde{\mathbf{h}}_m[n]$ denotes the interpolated CSI vector after delay-domain filtering, $\widetilde{\mathbf{h}}_{m,\mathrm{ref}}[n]$ denotes the background-referenced component, and $\odot$ denotes element-wise multiplication. Within each slow-time window, ReWiS then forms a minimum-variance distortionless-response (MVDR)-style range filter
\begin{equation}
\mathbf{w}_m(r)=\frac{\mathbf{R}_{xx,m}^{-1}\mathbf{s}(r)}{\mathbf{s}^{H}(r)\mathbf{R}_{xx,m}^{-1}\mathbf{s}(r)},
\end{equation}
where $\mathbf{s}(r)$ is the steering vector for candidate range $r$ and $\mathbf{R}_{xx,m}$ is the regularized covariance matrix of the motion-sensitive sequence. The resulting range-focused slow-time signal is converted to a Doppler-range response through a zoom fast Fourier transform (FFT),
\begin{equation}
D_m(\tau,\nu,r)=\left|\mathcal{F}_{\nu}\!\left\{\mathbf{w}_m^{H}(r)\mathbf{c}_m[n]\right\}_{n=\tau}^{\tau+L-1}\right|,
\end{equation}
where $\tau$ indexes the slow-time window, $\nu$ denotes the Doppler bin, $L$ is the slow-time window length, and $\mathcal{F}_{\nu}\{\cdot\}$ denotes the Doppler-axis Fourier transform. Let $R$ denote the number of retained range bins. The final per-antenna micro-Doppler map is then obtained by averaging over those $R$ bins,
\begin{equation}
\mathbf{a}_m(\tau,\nu)=\frac{1}{R}\sum_{r=1}^{R}D_m(\tau,\nu,r).
\end{equation}
The resulting sensing feature is therefore an explicit micro-Doppler map rather than an unspecified feature tensor.

\subsubsection{General Multichannel Stream Construction}
\begin{figure*}[!t]
    \centering
    \begin{minipage}[c]{0.45\textwidth}
        \centering
        \includegraphics[width=0.88\linewidth]{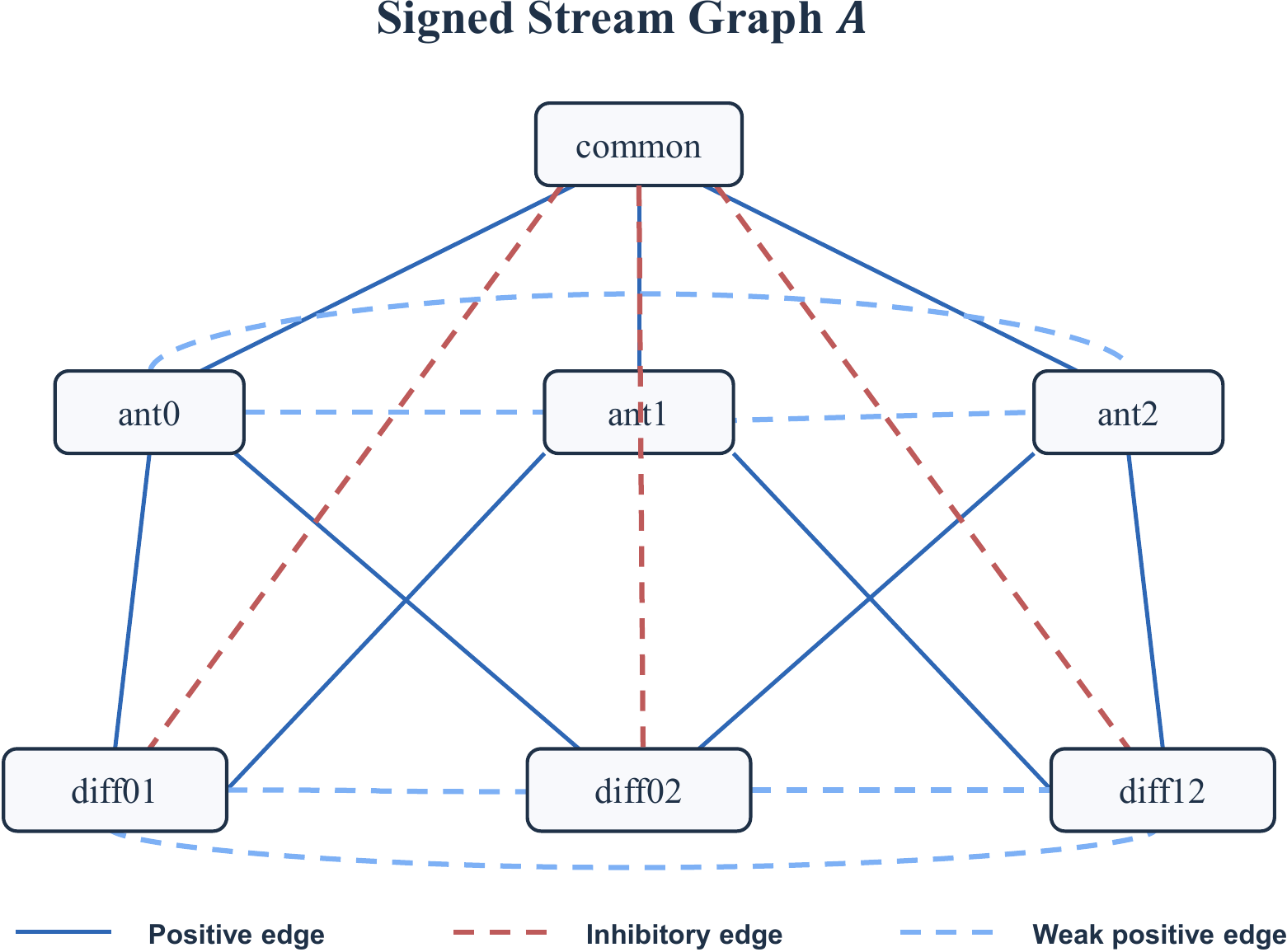}
       
        {\footnotesize (a) Signed stream graph}
    \end{minipage}\hfill
    \begin{minipage}[c]{0.55\textwidth}
        \centering
        \includegraphics[width=1.0\linewidth]{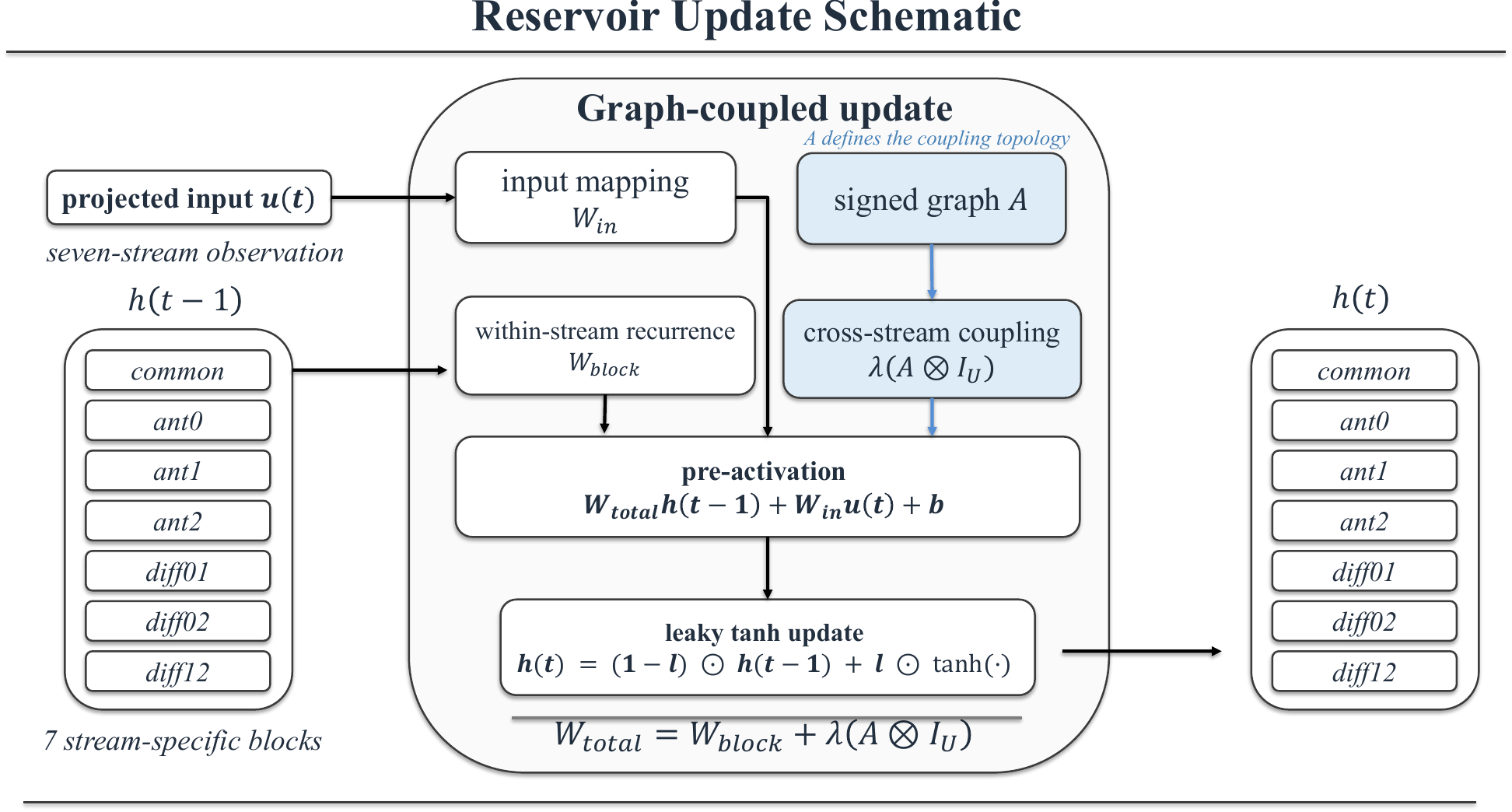}
       
        {\footnotesize (b) Reservoir update}
    \end{minipage}
    \caption{Reservoir-side structure in ReWiS for the Widar 3.0 with $M{=}3$ and $S{=}7$: (a) signed stream graph $\mathbf{A}$ over the seven structured streams; (b) reservoir update schematic. The projected streams remain the signal inputs, while $\mathbf{A}$ only specifies block-to-block coupling inside the shared recurrent operator.}
    \label{fig:reservoir_pair}
\end{figure*}

Let $M$ denote the number of receive channels. After the above preprocessing, the $m$-th receive channel is represented by one aligned micro-Doppler map
\begin{equation}
\mathbf{a}_m \in \mathbb{R}^{T \times F}, \qquad m=1,\dots,M,
\end{equation}
where $T$ and $F$ denote the temporal and Doppler resolutions of the sensing representation. ReWiS then reorganizes $\{\mathbf{a}_m\}_{m=1}^{M}$ into three complementary stream families. First, the common stream is
\begin{equation}
\mathbf{x}^{(c)}=\frac{1}{M}\sum_{m=1}^{M}\mathbf{a}_m.
\end{equation}
Second, the antenna-specific streams are
\begin{equation}
\mathbf{x}^{(a)}_m=\mathbf{a}_m,\qquad m=1,\dots,M.
\end{equation}
Third, the pairwise differential streams are
\begin{equation}
\mathbf{x}^{(d)}_{ij}=\frac{\mathbf{a}_i-\mathbf{a}_j}{\sqrt{2}}, \qquad 1\le i<j\le M.
\end{equation}
The resulting structured stream set is
\begin{equation}
\mathcal{X}=\left\{\mathbf{x}^{(c)}\right\}\cup\left\{\mathbf{x}^{(a)}_m\right\}_{m=1}^{M}\cup\left\{\mathbf{x}^{(d)}_{ij}\right\}_{1\le i<j\le M},
\end{equation}
with total stream count
\begin{equation}
S=1+M+\frac{M(M-1)}{2}.
\end{equation}
In the Widar 3.0 case used in our experiments, $M=3$ and therefore $S=7$, which recovers the seven streams shown in Figs.~\ref{fig:rewig_overview} and \ref{fig:reservoir_pair}. The three families have distinct physical meaning: $\mathbf{x}^{(c)}$ captures motion common to all antennas, $\mathbf{x}^{(a)}_m$ preserves view-dependent sensing responses, and $\mathbf{x}^{(d)}_{ij}$ emphasizes relative cross-antenna variations by suppressing common-mode content. This decomposition introduces physically meaningful inductive bias before temporal encoding rather than treating the WiFi observation as one undifferentiated tensor.

\subsection{Graph-Coupled Reservoir Encoding}
\subsubsection{Generalized Stream Graph Prior}
The reservoir keeps the structured WiFi streams as signal inputs and uses a signed graph only as a coupling prior among the recurrent blocks. Let the structured stream set be indexed as $\mathcal{S}=\{1,\dots,S\}$ and write $\mathbf{x}_s$ for the $s$-th stream in that order. ReWiS then defines a signed adjacency matrix
\begin{equation}
\mathbf{A}\in\mathbb{R}^{S\times S},
\end{equation}
whose signed entries encode whether common, antenna-specific, and differential streams should reinforce, weakly support, or suppress one another according to their relation. We write
\begin{equation}
A_{ij}=
\begin{cases}
1, & (i,j)\in\mathcal{E}_{\mathrm{main}}^{+},\\
\gamma_{\mathrm{neg}}, & (i,j)\in\mathcal{E}^{-},\\
\gamma_{\mathrm{intra}}, & (i,j)\in\mathcal{E}_{\mathrm{weak}}^{+},\\
0, & \text{otherwise},
\end{cases}
\end{equation}
where $\mathcal{E}_{\mathrm{main}}^{+}$ contains the main positive couplings, $\mathcal{E}^{-}$ contains inhibitory edges, and $\mathcal{E}_{\mathrm{weak}}^{+}$ contains weak same-group reinforcement edges. In physical terms, the common stream should reinforce antenna-specific motion views, suppress purely differential streams, and allow each differential stream to reconnect to the two antenna views from which it is formed. The graph is undirected in our implementation, so $\mathbf{A}$ is symmetric. Fig.~\ref{fig:reservoir_pair}(a) visualizes the Widar 3.0 instance of this prior for $M=3$ and $S=7$.

\subsubsection{Reservoir State Evolution}
Each structured stream $\mathbf{x}_s\in\mathbb{R}^{T\times F}$ is first mapped through a shared orthonormal random projection
\begin{equation}
\widetilde{\mathbf{x}}_s=\mathbf{x}_s\mathbf{P},\qquad \mathbf{P}\in\mathbb{R}^{F\times F_{\mathrm{proj}}},
\end{equation}
where $\mathbf{P}$ is shared across streams, orthonormalized once at initialization, and then kept fixed. At time step $t$, the projected stream features are concatenated into the actual reservoir input
\begin{equation}
\mathbf{u}_t=[\widetilde{\mathbf{x}}_{1,t};\widetilde{\mathbf{x}}_{2,t};\cdots;\widetilde{\mathbf{x}}_{S,t}]\in\mathbb{R}^{SF_{\mathrm{proj}}}.
\end{equation}
The WiFi signal is still the stream sequence $\{\mathbf{u}_t\}_{t=1}^{T}$; the graph only specifies how the recurrent blocks interact after those streams have entered the reservoir.

The reservoir allocates $U$ recurrent units to each stream block, giving the joint hidden state
\begin{equation}
\mathbf{h}_t=[\mathbf{h}_t^{(1)};\mathbf{h}_t^{(2)};\cdots;\mathbf{h}_t^{(S)}]\in\mathbb{R}^{SU},
\end{equation}
where each block $\mathbf{h}_t^{(s)}\in\mathbb{R}^{U}$ stores the temporal memory associated with stream $s$. The graph-coupled recurrent operator is
\begin{equation}
\mathbf{W}_{\mathrm{total}}=\mathbf{W}_{\mathrm{block}}+\lambda(\mathbf{A}\otimes \mathbf{I}_U),
\end{equation}
with $\mathbf{W}_{\mathrm{block}}=\mathrm{diag}(\mathbf{W}^{(1)},\dots,\mathbf{W}^{(S)})$, where $\lambda$ is a scalar graph-coupling coefficient and $\mathbf{I}_U$ is the $U\times U$ identity matrix. The diagonal blocks preserve within-stream recurrence, while the off-diagonal blocks introduced by $\lambda(\mathbf{A}\otimes \mathbf{I}_U)$ impose signed cross-stream interactions following Fig.~\ref{fig:reservoir_pair} (a).

Starting from $\mathbf{h}_0=\mathbf{0}$, the reservoir combines the recurrent drive $\mathbf{W}_{\mathrm{total}}\mathbf{h}_{t-1}$ with the input drive $\mathbf{W}_{\mathrm{in}}\mathbf{u}_t$, where
\begin{equation}
\mathbf{W}_{\mathrm{in}}\in\mathbb{R}^{SU\times SF_{\mathrm{proj}}}
\end{equation}
is a fixed random input-to-state matrix. Thus the previous state captures what the WiFi stream has accumulated over time, while the current projected observation injects the newest micro-Doppler evidence into the same block-structured state space. The update is
\begin{equation}
\mathbf{h}_t=(1-\boldsymbol{\ell})\odot \mathbf{h}_{t-1}+\boldsymbol{\ell}\odot \tanh\!\left(\mathbf{W}_{\mathrm{total}}\mathbf{h}_{t-1}+\mathbf{W}_{\mathrm{in}}\mathbf{u}_t+\mathbf{b}\right),
\end{equation}
where $\boldsymbol{\ell}$ is a heterogeneous leak-rate vector, $\mathbf{b}$ is a fixed bias vector, and $\odot$ denotes element-wise multiplication. After a short warm-up, the reservoir outputs
\begin{equation}
\mathbf{H}\in\mathbb{R}^{T'\times S\times U},
\end{equation}
where $T'$ denotes the temporal length after discarding the warm-up prefix. This output retains stream-wise temporal trajectories for downstream readout. Fig.~\ref{fig:reservoir_pair} (b) summarizes this flow: the WiFi observation forms the input-driven term, the previous hidden state contributes the recurrent drive, and the signed graph determines how the stream-specific recurrent blocks are coupled before the next hidden state is produced.

\subsection{Readout and Deployment Adaptation}
\subsubsection{Readout from Pooled Stream Descriptors}
After reservoir encoding, $\mathbf{H}$ contains one temporal state trajectory per structured stream. Temporal pyramid pooling compresses each trajectory into a fixed-length descriptor, using pooling levels $(1,2,4)$ and the statistics mean, standard deviation, and mean absolute temporal difference. Stacking the pooled stream descriptors yields
\begin{equation}
\mathbf{Z}\in\mathbb{R}^{S\times D_{\mathrm{pool}}},
\end{equation}
where row $s$ summarizes the temporal dynamics of stream $s$ and $D_{\mathrm{pool}}$ denotes the pooled descriptor dimension induced by the selected pooling levels and statistics. The reservoir thus converts each WiFi stream into a state trajectory, and pooling reduces each trajectory to one compact stream descriptor. The readout treats the $S$ rows of $\mathbf{Z}$ as graph nodes and lets them exchange information with the same signed prior that was used on the reservoir side. Let $\widetilde{\mathbf{A}}=\mathbf{A}+\mathbf{I}_{S}$, where $\mathbf{I}_{S}$ is the $S\times S$ identity matrix, and let $\mathbf{D}$ be the diagonal matrix with $D_{ii}=\sum_j |\widetilde{A}_{ij}|$. The normalized signed adjacency is
\begin{equation}
\widehat{\mathbf{A}}=\mathbf{D}^{-\frac{1}{2}}\widetilde{\mathbf{A}}\mathbf{D}^{-\frac{1}{2}}.
\end{equation}
If $\mathbf{R}^{(0)}$ denotes the node-encoded feature matrix obtained from $\mathbf{Z}$ by the lightweight node encoder, each residual graph-convolution stage is
\begin{equation}
\mathbf{R}^{(\ell+1)}=\mathbf{R}^{(\ell)}+\Phi\!\left(\widehat{\mathbf{A}}\mathbf{R}^{(\ell)}\right),
\end{equation}
where $\Phi(\cdot)$ denotes a lightweight learnable transform. In WiFi terms, this stage allows common, antenna-specific, and differential stream evidence to interact again after temporal encoding, rather than forcing the final decision to rely on each stream independently.

The updated node features are then merged into one graph-level descriptor by attention pooling. Let $\mathbf{r}_s$ be the final node feature of stream $s$. The attention weights and graph-level feature are
\begin{equation}
\alpha_s=\frac{\exp(q(\mathbf{r}_s))}{\sum_{j=1}^{S}\exp(q(\mathbf{r}_j))}, \qquad
\mathbf{g}=\sum_{s=1}^{S}\alpha_s\mathbf{r}_s,
\end{equation}
where $q(\cdot)$ is a scalar attention-scoring function. The final prediction is produced by
\begin{equation}
\widehat{\mathbf{y}}=f_{\mathrm{cls}}(\mathbf{g}).
\end{equation}
Thus the readout first converts reservoir states into pooled stream descriptors, then performs signed graph interaction among those descriptors, and finally forms an attention-weighted global descriptor for the sensing output $\widehat{\mathbf{y}}$.

\subsubsection{Readout-Only Deployment Adaptation}
After a WiFi sensing node is deployed in a new room or under a new user condition, we assume that only a small $K$-shot target set
\begin{equation}
\mathcal{D}_{\mathrm{tgt}}^{(K)}=\{(\mathcal{X}_n,y_n)\}_{n=1}^{N_K}
\end{equation}
is available from the new domain, where $K$ denotes the number of labeled target examples per class, $C$ is the number of classes, and $N_K=KC$ is the total number of labeled target samples. For deployment, ReWiS keeps the reservoir-side parameters
\begin{equation}
\Theta_{\mathrm{res}}=\{\mathbf{P},\mathbf{W}_{\mathrm{total}},\mathbf{W}_{\mathrm{in}},\mathbf{b},\boldsymbol{\ell}\}
\end{equation}
fixed and updates only the lightweight readout parameters
\begin{equation}
\Theta_{\mathrm{ro}}=\{\Theta_{\mathrm{enc}},\Theta_{\mathrm{gr}},\Theta_{\mathrm{att}},\Theta_{\mathrm{cls}}\},
\end{equation}
where $\Theta_{\mathrm{enc}}$, $\Theta_{\mathrm{gr}}$, $\Theta_{\mathrm{att}}$, and $\Theta_{\mathrm{cls}}$ denote the node encoder, graph readout, attention pooler, and classifier parameters, respectively. Adaptation starts from the source-trained readout rather than from random initialization. The stream construction and reservoir-side temporal encoding remain unchanged after installation, and only the lightweight decision layers are retuned to the new room, user, or receiver condition.

The adaptation objective is
\begin{equation}
\begin{aligned}
\Theta_{\mathrm{ro}}^{\star}=\arg\min_{\Theta_{\mathrm{ro}}}\ &
\frac{1}{N_K}\sum_{n=1}^{N_K}\mathcal{L}_{\mathrm{CE}}\!\left(f(\mathcal{X}_n;\Theta_{\mathrm{res}},\Theta_{\mathrm{ro}}),y_n\right)\\
&+\beta\,\Omega(\Theta_{\mathrm{ro}},\Theta_{\mathrm{ro}}^{\mathrm{src}}),
\end{aligned}
\end{equation}
where $f(\mathcal{X}_n;\Theta_{\mathrm{res}},\Theta_{\mathrm{ro}})$ denotes the prediction function of ReWiS, $\mathcal{L}_{\mathrm{CE}}$ is the cross-entropy loss, $\Theta_{\mathrm{ro}}^{\mathrm{src}}$ denotes the source-trained readout parameters, $\beta$ is the regularization weight, and $\Omega$ is a regularizer that discourages large deviations from the source initialization when $K$ is small. This setting keeps post-deployment updates on the smallest trainable portion of the model, avoids re-optimizing the reservoir-side temporal encoder, and preserves the source-domain sensing prior while still allowing recovery on the new domain.

\section{Experimental Evaluation}
\subsection{Dataset and evaluation protocol}

We evaluate ReWiS on Widar 3.0 \cite{zhang2022widar3}, a large WiFi sensing benchmark containing 271,013 samples from 22 gesture classes. Representative gestures include Clap, Push\&Pull, Sweep, Slide, Draw-O, and Draw-Zigzag, while the full release spans a much broader gesture inventory. The dataset covers 17 users, 3 rooms, 6 receiver settings, 5 face-orientation settings, and 8 torso-location settings, making it a suitable platform for validating RC under realistic cross-domain WiFi sensing variation rather than only within one fixed gesture setup. We report one in-domain split and five cross-domain factors, namely room, user identity, receiver position, face orientation, and torso location. For each factor, we follow a leave-one-value-out evaluation protocol, so one factor value is treated as the target domain and the remaining values are used for source-domain training. For deployment adaptation, we sample $K\in\{1,3,5,10,20,50,100,200\}$ labeled target examples per class and update only the readout head.

\subsection{Baselines and Metrics}

We compare ReWiS with three lightweight-to-moderate deep baselines trained under the same protocol: MobileViT-XXS\cite{mehta2022mobilevit}, EfficientNet-B0\cite{tan2019efficientnet}, and ResNet18\cite{he2016resnet}. These baselines were not chosen arbitrarily. MobileViT-XXS represents a mobile-scale hybrid CNN-transformer model, EfficientNet-B0 represents a compact convolutional backbone with strong parameter efficiency, and ResNet18 represents a widely used residual CNN with moderate capacity and stable optimization behavior. They span practical deployment baselines of increasing model complexity while remaining feasible on commodity WiFi-sensing platforms. All models use the same CSI-derived micro-Doppler inputs from Section~II. We report macro-F1 as the main recognition metric and additionally compare trainable parameter count, time to the best checkpoint, and single-sample CPU latency.

\subsection{Implementation Details}

ReWiS uses projection dimension 48, 96 reservoir units per stream, coupling strength 0.14, warm-up length 6, heterogeneous leak rates 0.22 and 0.62, signed-graph weights $\gamma_{\mathrm{neg}}=-0.8$ and $\gamma_{\mathrm{intra}}=0.3$, and temporal pyramid pooling levels $(1,2,4)$ with mean, standard deviation, and mean absolute temporal difference. For latency measurement, we use batch size 1 with 10 warm-up runs and 40 timed repeats on a workstation equipped with a 12th Gen Intel(R) Core(TM) i5-12600KF CPU.

\vspace{-0.3em}

\section{Results and Discussion}
\subsection{Overall Recognition Performance}
\begin{table}[t]
\centering
\caption{Macro-F1 (\%) on Widar 3.0.}
\label{tab:main_results}
\scriptsize
\setlength{\tabcolsep}{3pt}
\renewcommand{\arraystretch}{1.05}
\resizebox{\columnwidth}{!}{%
\begin{tabular}{lccccccc}
\toprule
Model & In-dom. & Room & Identity & Receiver & Face & Torso & Cross-dom. mean \\
\midrule
MViT-XXS & 94.8 & 83.2 & 86.3 & 91.1 & 89.7 & 87.0 & 87.5 \\
EffNet-B0 & 95.5 & 83.5 & 86.9 & 91.9 & 90.2 & 87.3 & 88.0 \\
ResNet18 & \textbf{96.3} & \textbf{85.2} & \textbf{88.2} & \textbf{92.3} & \textbf{91.6} & \textbf{88.4} & \textbf{89.2} \\
\textbf{ReWiS} & 89.2 & 77.7 & 81.2 & 83.8 & 84.3 & 83.0 & 82.0 \\
\bottomrule
\end{tabular}
}
\end{table}

Table~\ref{tab:main_results} summarizes recognition performance under one in-domain split and five cross-domain transfer settings. Although the heavier deep baselines achieve higher peak macro-F1, the gap remains limited while the training burden of ReWiS is much smaller because only the lightweight readout is optimized. Its mean cross-domain macro-F1 is 82.0\%, compared with 87.5\% for MobileViT-XXS, 88.0\% for EfficientNet-B0, and 89.2\% for ResNet18, and it remains strong on face orientation (84.3\%) and receiver transfer (83.8\%). In other words, the reservoir-centric pipeline sacrifices only a moderate amount of macro-F1 while keeping the sensing model much lighter and easier to adapt.

\begin{figure*}[t]
\centering
\begin{minipage}[t]{0.339\textwidth}
\vspace{0pt}
\centering
\includegraphics[width=1.0\linewidth]{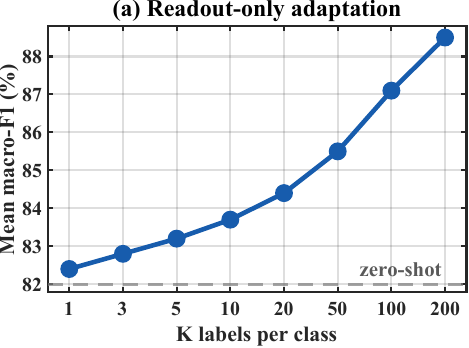}

\end{minipage}%
\begin{minipage}[t]{0.322\textwidth}
\vspace{0pt}
\centering
\includegraphics[width=1.0\linewidth]{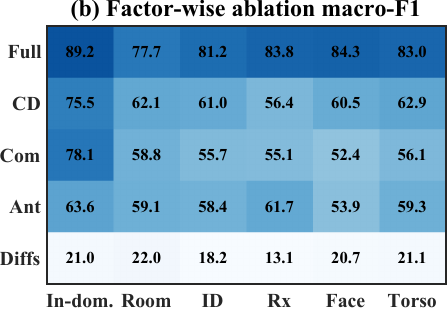}

\end{minipage}%
\begin{minipage}[t]{0.339\textwidth}
\vspace{0pt}
\centering
\includegraphics[width=1.0\linewidth]{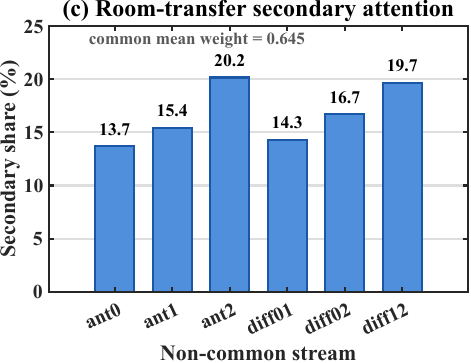}

\end{minipage}
\caption{Results beyond overall macro-F1. (a) Readout-only deployment adaptation with the reservoir frozen. (b) Factor-wise ablation across one in-domain split and five cross-domain factors. (c) Room-transfer secondary attention after excluding the dominant ``common'' stream.}
\label{fig:exp23_results}
\end{figure*}

\subsection{Readout-Only Deployment Adaptation}
We next evaluate readout-only deployment adaptation by freezing the reservoir and varying the number of labeled target examples per class from $K{=}1$ to $K{=}200$. Fig.~\ref{fig:exp23_results} (a) shows a consistent upward trend in mean macro-F1: averaged over the five domain factors, ReWiS improves from 82.0\% in the zero-shot setting to 84.4\% with only $K{=}20$ labeled target samples and to 88.5\% with $K{=}200$. The gains are especially clear on the harder room and identity shifts, where the model improves from 77.7\% to 82.8\% and from 81.2\% to 90.4\%, respectively. This result is important from the WiFi deployment viewpoint: once a small number of target labels becomes available after installation, the sensing model can recover much of the new-domain loss without touching the reservoir-side temporal encoder.

\subsection{Effect of the Structured Seven-Stream Input}
Figs.~\ref{fig:exp23_results} (b) and \ref{fig:exp23_results} (c) together test whether the structured seven-stream decomposition is actually needed. The full seven-stream model achieves the best macro-F1 in every evaluated domain factor, reaching 89.2\% in-domain and 77.7\%, 81.2\%, 83.8\%, 84.3\%, and 83.0\% on room, identity, receiver, face-orientation, and torso transfer, respectively. All reduced variants drop markedly, with Diffs-only performing worst across all settings and Common+Diffs still leaving a clear gap to the full model. The attention view in Fig.~\ref{fig:exp23_results} (c) is consistent with this pattern: after excluding the dominant "common" stream, which aggregates motion shared across antennas, the largest remaining weights fall on ant2 and the pairwise differential streams diff12 and diff02. Together, this indicates that relative cross-antenna motion cues still contribute materially to the final sensing decision.

\vspace{-0.5em}

\subsection{Efficiency and Deployment Cost}
\begin{table}[t]
\centering
\caption{Efficiency on the representative room-transfer split.}
\label{tab:efficiency}
\scriptsize
\setlength{\tabcolsep}{4pt}
\renewcommand{\arraystretch}{1.05}
\resizebox{\columnwidth}{!}{%
\begin{tabular}{lccc}
\toprule
Model & Params (M) & Room $t_{\mathrm{best}}$ (min) & CPU latency (ms/sample) \\
\midrule

MViT-XXS & 1.06 & 63 & 9.17 \\
EffNet-B0 & 4.36 & 87 & 11.61 \\
ResNet18 & 11.34 & 137 & 11.78 \\
\textbf{ReWiS} & \textbf{0.72} & \textbf{6} & \textbf{6.24} \\
\bottomrule
\end{tabular}
}
\end{table}

Table~\ref{tab:efficiency} highlights one of ReWiS's defining practical advantages: the reservoir itself is never trained, so both source-domain fitting and post-deployment retuning reduce to optimizing a compact readout rather than repeatedly updating a deep temporal encoder. This difference is already clear on the representative room split. Averaged over the three leave-one-room-out room folds on a workstation equipped with a 12th Gen Intel(R) Core(TM) i5-12600KF CPU, ReWiS reaches its best validation checkpoint in about 6 minutes and 22 epochs, whereas MobileViT-XXS, EfficientNet-B0, and ResNet18 require about 63 minutes and 10 epochs, 87 minutes and 10 epochs, and 137 minutes and 19 epochs, respectively. Under the same cached seven-stream input boundary, ReWiS also achieves the lowest single-sample CPU latency at 6.24\,ms/sample, compared with 9.17, 11.61, and 11.78\,ms/sample. In short, ReWiS is not only lighter at inference time; its fixed-reservoir design makes training and later readout-only adaptation substantially cheaper, which is an advantage for deployable WiFi sensing.

\vspace{-0.5em}

\section{Conclusion}

We presented ReWiS, a deployment-oriented reservoir framework for cross-domain WiFi sensing. By combining structured micro-Doppler stream construction, fixed graph-coupled reservoir dynamics, and readout-only deployment adaptation, ReWiS preserves temporal modeling while keeping optimization and post-deployment updates lightweight. On Widar 3.0, ReWiS reaches 82.0\% mean cross-domain macro-F1, improves to 88.5\% under limited target supervision, and remains cheaper than heavier deep baselines in trainable parameters, optimization time, and single-sample CPU latency. More important, ReWiS points to a practical WiFi sensing pipeline in which sensing-oriented temporal encoding remains stable on deployed AP/router-class nodes while only a small output head is updated online, and it is also compatible with future low-power WiFi sensing hardware, where analog, photonic, memristive, or related physical reservoirs may serve as front-end temporal encoders for edge-side sensing nodes.

\vspace{-0.7em}

\bibliographystyle{IEEEtran}
\bibliography{refs}

\end{document}